\documentclass[aps,preprint]{revtex4}%
\usepackage{amsfonts}
\usepackage{amsmath}
\usepackage{amssymb}
\usepackage{graphicx}%
\providecommand{\U}[1]{\protect\rule{.1in}{.1in}}

\begin{document}
\author{S. Bruce}
\affiliation{Department of Physics, University of Concepcion, P.O. Box 160C, Concepcion, Chile}
\title{The Schr\"{o}dinger equation and negative energies}
\date{06/30/08}

\begin{abstract}
We present a nonrelativistic wave equation for the electron in $\left(
3+1\right)  $-dimensions which includes negative-energy eigenstates. We solve
this equation for three well-known instances, reobtaining the corresponding
Pauli equation (but including negative-energy eigenstates) in each case.

PACS 03.65.Bz - Foundations, theory of measurements, miscellaneous theories.

PACS 03.65.Pm - Relativistic wave equations.

\end{abstract}
\maketitle

\section{Introduction}

Let $K$ and $K^{\prime}$ be two inertial reference frames. The $K^{\prime}$
system has all its axes parallel to those of $K$ and moves with velocity
$\mathbf{v}$ relative to $K.$ Assume $E,$ $\mathbf{p}$ and $E^{\prime},$
$\mathbf{p}^{\prime}$ to be the energy and momentum of a particle of mass
$m_{0}$ in the $K$ and $K^{\prime}$ frames, respectively. The Lorentz
transformation for these quantities is \cite{JA}
\begin{equation}
\left\{
\begin{array}
[c]{c}%
E=\mathbf{v\cdot p+}E^{\prime}\gamma^{-1}\ ,\qquad\qquad\quad\quad\quad
\quad\ \ \ \ \\
\mathbf{p}=\mathbf{p}^{\prime}\mathbf{+}\left(  \gamma-1\right)
\mathbf{v}\left(  \mathbf{p}^{\prime}\cdot\mathbf{v}\right)  \mathbf{/v}%
^{2}\mathbf{+}\gamma\mathbf{v}E^{\prime}/c^{2},
\end{array}
\right.  \label{02}%
\end{equation}
with $\gamma=\left(  1-\mathbf{v}^{2}/c^{2}\right)  ^{-1/2}$, which is also
valid for eventual negative values of $E$ and $E^{\prime}$. Let us now assume
that the particle is at rest in the $K^{\prime}$ system, i.e., $E^{\prime
}=m_{0}c^{2}$. Thus the first Eq.(\ref{02}) becomes
\begin{equation}
E=\mathbf{v\cdot\mathbf{p}+}m_{0}c^{2}\ \gamma^{-1}\ . \label{03}%
\end{equation}
It is interesting to note that this expression represents a
\textit{correspondence relation} for $E_{D}$: the energy eigenvalues of the
free-particle Dirac Hamiltonian. To see this, let us calculate the eigenvalues
of the Dirac Hamiltonian \cite{MA} in the energy-helicity eigenstates $\mid
E_{D},s\mathbf{\rangle}$ :
\begin{equation}
\widehat{H}_{D}\mid E_{D},s\mathbf{\rangle}=\left(  c\mathbf{\alpha
\cdot\widehat{\mathbf{p}}}+m_{0}c^{2}\beta\right)  \mid E_{D},s\mathbf{\rangle
=}E_{D}\mid E_{D},s\mathbf{\rangle}. \label{033}%
\end{equation}
To simplify our notation, let $\widehat{\mathcal{O}}$ be a matrix operator.
Define $\langle\widehat{\mathcal{O}}\mathbf{\rangle}_{s}\equiv\langle
E_{D},s\mid\widehat{\mathcal{O}}\mid E_{D},s\mathbf{\rangle}$\textbf{.} Thus%
\begin{equation}
E_{D}\equiv\langle\widehat{H}_{D}\mathbf{\rangle}_{s}\mathbf{=}\langle
c\mathbf{\alpha\rangle}_{s}\mathbf{\cdot p}+\langle\beta\mathbf{\rangle}%
_{s}\mathbf{\ }m_{0}c^{2}\ . \label{04}%
\end{equation}
Therefore%
\begin{equation}
\langle\left\{  c\mathbf{\alpha\ ,\ }\widehat{H}_{D}\right\}  \mathbf{\rangle
}_{s}\mathbf{=\ }c^{2}\mathbf{p\ ,\qquad}\langle\left\{  \beta\ \mathbf{,\ }%
\widehat{H}_{D}\right\}  \mathbf{\rangle}_{s}\mathbf{=\ }m_{0}c^{2}\ ,
\end{equation}
where the symbol $\left\{  \ \mathbf{\ ,\ \ }\right\}  $ means \textit{anti}%
commutation relation. By using $\alpha^{\prime}s$ and $\beta$ matrices
properties%
\begin{equation}
\left\{  \alpha_{i}\ ,\ \alpha_{j}\right\}  =2\delta_{ij},\qquad\left\{
\beta\ ,\ \alpha_{j}\right\}  =0,\qquad\beta^{2}=I,\qquad i,j=1,2,3,
\label{06}%
\end{equation}
we get%
\begin{equation}
\left\langle c\mathbf{\alpha}\right\rangle _{s}\mathbf{=}\frac{c^{2}%
\mathbf{p}}{E_{D}}\mathbf{\ ,\qquad}\left\langle \beta\right\rangle
_{s}\mathbf{=}\frac{M_{0}c^{2}}{E_{D}}=\gamma^{-1}\ .
\end{equation}
Hence, from Eq.(\ref{03}) and Eq.(\ref{04}), we make the following
association
\begin{equation}
\left\langle \widehat{H}_{D}\right\rangle _{s}=E_{D}\rightarrow E,\qquad
\left\langle c\mathbf{\alpha}\right\rangle _{s}\rightarrow\mathbf{v,\qquad
}\left\langle \beta\right\rangle _{s}\rightarrow\gamma^{-1}.
\end{equation}
Notice that a separation of concepts is involved in Eqs.(\ref{03}) and
(\ref{04}). The term $\mathbf{v\cdot p}$ in Eq.(\ref{03}) is the scalar
product of the velocity $\mathbf{v}$ of the frame $K^{\prime}$ (where the
particle is at rest) and the momentum $\mathbf{p}$ of the particle, both
relative to the (laboratory) frame $K$. They are associated with different
objects. Therefore, at the quantum level, we have to distinguish between
$\mathbf{p}/m_{0}$ and $\gamma\mathbf{v}$\textbf{,} although they have the
same \textit{mean} value. Then, \textit{loosely} speaking, the Dirac
Hamiltonian is \textit{formally }linked with a Lorentz transformation itself.

\section{Description of a nonrelativistic spin-1/2 particle}

To begin with, for a \textit{strictly} nonrelativistic wave equation
($\left\vert \mathbf{v}\right\vert \ll c$), we are concerned with the
expression%
\begin{equation}
E=m_{0}c^{2}+\frac{\mathbf{p}^{2}}{2m_{0}}\ . \label{10}%
\end{equation}
One alternative of (limited) \textit{linearization }of\textit{\ }this equation
is done by first squaring Eq.(\ref{10}):%
\begin{equation}
E^{2}=\left(  m_{0}c^{2}+\frac{\mathbf{p}^{2}}{2m_{0}}\right)  ^{2}%
=c^{2}\mathbf{p}^{2}+m_{0}^{2}c^{4}+\frac{\left(  \mathbf{p}^{2}\right)  ^{2}%
}{4m_{0}^{2}}\ . \label{12}%
\end{equation}
Then, taking the \textquotedblleft square root\textquotedblright\ of
Eq.(\ref{12}), in the same way as the Dirac Hamiltonian is discerned from the
Klein-Gordon equation, we find the wave equation \cite{BR}
\begin{equation}
\widehat{H}\Phi\left(  \mathbf{r,}t\right)  =\left(  c\mathbf{\alpha}%
\cdot\mathbf{\widehat{\mathbf{p}}}+m_{0}c^{2}\beta+i\beta\gamma_{5}%
\frac{\left(  \mathbf{\alpha}\cdot\mathbf{\widehat{\mathbf{p}}}\right)  ^{2}%
}{2m_{0}}\right)  \Phi\left(  \mathbf{r,}t\right)  =i\hbar\frac{\partial
}{\partial t}\Phi\left(  \mathbf{r,}t\right)  . \label{20}%
\end{equation}
This resembles a single-particle wave equation with Hamiltonian%
\begin{equation}
\widehat{H}=c\mathbf{\alpha}\cdot\mathbf{\widehat{\mathbf{p}}}+m_{0}c^{2}%
\beta+i\beta\gamma_{5}\frac{\left(  \mathbf{\alpha}\cdot\mathbf{\widehat
{\mathbf{p}}}\right)  ^{2}}{2m_{0}}\ .
\end{equation}
Here
\begin{equation}
\gamma_{k}=-i\beta\alpha_{k},\qquad\gamma_{0}=\beta,\qquad\gamma_{5}%
\equiv\gamma_{1}\gamma_{2}\gamma_{3}\gamma_{0},\qquad\gamma_{\mu,5}^{\dag
}=\gamma_{\mu,5}\ ,
\end{equation}
where we use the metric signature $g(+---)$. Thus the wave function
$\Phi\left(  \mathbf{r,}t\right)  $ is (in Eq.(\ref{20})) a bispinor. Note
that Eq.(\ref{20}) introduces the degree of freedom for nonrelativistic
antiparticles. In this way we have incorporated \textit{negative-}energy
eigenvalues in the nonrelativistic limit, so that the electron
\textit{Zitterbewegung} is also present.

Now we mimic the Dirac case in order to calculate mean values:%
\begin{equation}
\langle\left\{  \widehat{H},\mathbf{\alpha}\right\}  \mathbf{\rangle}%
_{s}\mathbf{=\ }2c\mathbf{p\ },\qquad\langle\left\{  \widehat{H}%
,\beta\right\}  \mathbf{\rangle}_{s}=2m_{0}c\ ,\qquad\langle\left\{
\widehat{H},i\beta\gamma_{5}\right\}  \mathbf{\rangle}_{s}=\frac
{\mathbf{p}^{2}}{m_{0}}\ .
\end{equation}
From these equations we get%
\begin{equation}
\left\langle \mathbf{\alpha}\right\rangle _{s}\mathbf{=}\frac{c\mathbf{p}}%
{E}\ ,\qquad\left\langle \beta\right\rangle _{s}=\frac{m_{0}c^{2}}{E}%
\ ,\qquad\left\langle i\beta\gamma_{5}\right\rangle _{s}\mathbf{=}%
\frac{\mathbf{p}^{2}}{2m_{0}E}\ .
\end{equation}

Next we regard a nonrelativistic electron moving in the presence of a
classical magnetic field. To establish the wave equation governing this
system, we make the minimal replacement%
\begin{equation}
\widehat{\mathbf{p}}\rightarrow\widehat{\mathbf{\Pi}}=\widehat{\mathbf{p}%
}-\frac{e}{c}\mathbf{A}(\mathbf{r,}t) \label{40}%
\end{equation}
in Eq.(\ref{20}), obtaining%
\begin{equation}
\widehat{H}\Phi\left(  \mathbf{r,}t\right)  =\left\{  c\mathbf{\alpha}%
\cdot\widehat{\mathbf{\Pi}}+m_{0}c^{2}\beta+\frac{i}{2m_{0}}\beta\gamma
_{5}\left(  \mathbf{\alpha}\cdot\widehat{\mathbf{\Pi}}\right)  ^{2}\right\}
\Phi\left(  \mathbf{r,}t\right)  =i\hbar\frac{\partial}{\partial t}\Phi\left(
\mathbf{r,}t\right)  . \label{30}%
\end{equation}
Applying $\widehat{H}$ on the left hand side of Eq.(\ref{30}), we get%
\begin{align}
\widehat{H}^{2}\Phi\left(  \mathbf{r,}t\right)   &  =\left\{  \left(
c\mathbf{\alpha}\cdot\widehat{\mathbf{\Pi}}\right)  ^{2}+m_{0}^{2}c^{4}%
+\frac{1}{4m_{0}^{2}}\left(  \left(  \mathbf{\alpha}\cdot\widehat{\mathbf{\Pi
}}\right)  ^{2}\right)  ^{2}\right\}  \Phi\left(  \mathbf{r,}t\right)
\label{31}\\
&  =\left(  m_{0}c^{2}+\frac{1}{2m_{0}}\left(  \mathbf{\alpha}\cdot
\widehat{\mathbf{\Pi}}\right)  ^{2}\right)  \left(  m_{0}c^{2}+\frac{1}%
{2m_{0}}\left(  \mathbf{\alpha}\cdot\widehat{\mathbf{\Pi}}\right)
^{2}\right)  \Phi\left(  \mathbf{r,}t\right)  =-\hbar^{2}\frac{\partial^{2}%
}{\partial t^{2}}\Phi\left(  \mathbf{r,}t\right)  .\nonumber
\end{align}
From Eq.(\ref{31}), the wave equation that determines the $\Phi\left(
\mathbf{r,}t\right)  $ states is
\begin{equation}
\widehat{H}\Phi\left(  \mathbf{r,}t\right)  =\left(  m_{0}c^{2}+\frac
{1}{2m_{0}}\left(  \mathbf{\alpha}\cdot\widehat{\mathbf{\Pi}}\right)
^{2}\right)  \Phi\left(  \mathbf{r,}t\right)  =i\hbar\frac{\partial}{\partial
t}\Phi\left(  \mathbf{r,}t\right)  . \label{32}%
\end{equation}
The upper and lower components of $\Phi$ fulfil, up to a minus (-) sign (for
the negative-energy eigenstates), the same wave equation. A general solution
to Eq.(\ref{32}) is a linear combination of the eigenstates%
\begin{equation}
\Phi_{\pm}\left(  \mathbf{r},t\right)  =\Phi\left(  \mathbf{r}\right)
\exp\left(  -\frac{i}{\hbar}E_{\pm}t\right)  , \label{50}%
\end{equation}
with $E_{\pm}=\pm\left\vert E\right\vert $.

Finally, we address the case of a nonrelativistic spin-1/2 particle moving in
the presence of a scalar central potential. Equation (\ref{30}) can be written
as%
\begin{equation}
\frac{\mathbf{p}^{2}}{2m_{0}}=E_{\pm}-m_{0}c^{2}\ .
\end{equation}
For the free particle, Eq.(\ref{20}) turns out to be%
\begin{equation}
E_{\pm}\Phi\left(  \mathbf{r}\right)  =\left(  c\mathbf{\alpha}\cdot
\mathbf{\widehat{\mathbf{p}}}+m_{0}c^{2}\beta+i\beta\gamma_{5}\left(  E_{\pm
}-m_{0}c^{2}\right)  \right)  \Phi\left(  \mathbf{r}\right)  .
\end{equation}
Reordering the terms we find that%
\begin{equation}
E_{\pm}\Gamma_{1}\Phi\left(  \mathbf{r}\right)  =\left(  c\mathbf{\alpha}%
\cdot\mathbf{\widehat{\mathbf{p}}}+m_{0}c^{2}\Gamma_{2}\right)  \Phi\left(
\mathbf{r}\right)  , \label{60}%
\end{equation}
where $\Gamma_{1,2}$ are Hermitian operators defined as%
\begin{equation}
\Gamma_{1}\equiv\left(  I-i\beta\gamma_{5}\right)  ,\qquad\Gamma_{2}%
\equiv\left(  I+i\gamma_{5}\right)  \beta, \label{63}%
\end{equation}
with the properties%
\begin{equation}
\Gamma_{1}^{2}=2\Gamma_{1},\qquad\Gamma_{2}^{2}=2I,\qquad\Gamma_{2}\Gamma
_{1}=\left(  I+\beta\right)  \left(  I-i\gamma_{5}\right)  . \label{65}%
\end{equation}
Equation (\ref{60}) is now linear in $E_{\pm}$ and $\mathbf{p}$. It stands for
a generalization of L\'{e}vy-Leblond's wave equation \cite{LE,HA,HA2} since it
also includes \textit{negative}-energy eigenstates. Notice that $\Gamma_{1}/2$
is a \textit{singular} (projection) matrix. Hence, it does not have an
\textit{inverse} operator. This means that, unlike Eq.(\ref{20}), it is
\textit{not} possible to define an appropriate Hamiltonian from Eq.(\ref{60}):
expression (\ref{60}) has to be treated just as a \textit{wave equation}.
Therefore a nonrelativistic electron moving in the presence of a central
potential $V\left(  r\right)  $ satisfies
\begin{equation}
\left(  E_{\pm}-eV\left(  r\right)  \right)  \Gamma_{1}\Phi\left(
\mathbf{r}\right)  =\left(  c\mathbf{\alpha}\cdot\mathbf{\widehat{\mathbf{p}}%
}+m_{0}c^{2}\Gamma_{2}\right)  \Phi\left(  \mathbf{r}\right)  . \label{70}%
\end{equation}
Subsequently, multiplying Eq.(\ref{70}) on the left by $\Gamma_{2}$, and using
the fact that $\left(  I+i\gamma_{5}\right)  \left(  I-i\gamma_{5}\right)
/2=I$, yields
\begin{equation}
\left(  E_{\pm}-eV\left(  r\right)  \right)  \Gamma_{2}\Gamma_{1}\Phi\left(
\mathbf{r}\right)  =\left(  \Gamma_{2}c\mathbf{\alpha}\cdot\mathbf{\widehat
{\mathbf{p}}}+2m_{0}c^{2}\right)  \frac{1}{2}\Gamma_{2}\beta\left(
I-i\gamma_{5}\right)  \Phi\left(  \mathbf{r}\right)  .
\end{equation}
Given the fact that $\Gamma_{2}^{2}=2I$ and $\Gamma_{2}\beta=\left(
I+i\gamma_{5}\right)  $, it follows
\begin{equation}
\left(  E_{\pm}-eV\left(  r\right)  \right)  \left(  I+\beta\right)  \left[
\beta\left(  I-i\gamma_{5}\right)  \Phi\left(  \mathbf{r}\right)  \right]
=\left(  \Gamma_{2}c\mathbf{\alpha}\cdot\mathbf{\widehat{\mathbf{p}}}%
\Gamma_{2}+2m_{0}c^{2}\Gamma_{2}\right)  \frac{1}{2}\left[  \beta\left(
I-i\gamma_{5}\right)  \Phi\left(  \mathbf{r}\right)  \right]  , \label{71}%
\end{equation}
where we have used the property $\left(  I+\beta\right)  \beta=\left(
I+\beta\right)  $ on the left hand side of Eq.(\ref{71}). Next we define the
bispinor
\begin{equation}
\Psi\left(  \mathbf{r}\right)  \equiv\Gamma_{2}\Phi\left(  \mathbf{r}\right)
=\left(
\begin{array}
[c]{c}%
\phi\left(  \mathbf{r}\right) \\
\chi\left(  \mathbf{r}\right)
\end{array}
\right)  . \label{80}%
\end{equation}
Thus the wave function $\Psi$ corresponds to a rotation of (the bispinor)
$\Phi$ in the Hilbert space. This ensues from Eqs.(\ref{63}) and (\ref{65}),
since $\Gamma_{2}$ is both unitary and Hermitian. Thus%
\begin{equation}
\left(  E_{\pm}-eV\left(  r\right)  \right)  \left(  I+\beta\right)
\Psi\left(  \mathbf{r}\right)  =\left(  \frac{1}{2}\Gamma_{2}c\mathbf{\alpha
}\cdot\mathbf{\widehat{\mathbf{p}}}\Gamma_{2}+m_{0}c^{2}\Gamma_{2}\right)
\Psi\left(  \mathbf{r}\right)  .
\end{equation}
From the first Eq.(\ref{65}) and the property $\left\{  \mathbf{\alpha}%
\cdot\mathbf{\widehat{\mathbf{p}}\ ,\ }\Gamma_{2}\right\}  =0$, it follows
that
\begin{equation}
\left(  E_{\pm}-eV\left(  r\right)  \right)  \left(  I+\beta\right)
\Psi\left(  \mathbf{r}\right)  =\left(  -c\mathbf{\alpha}\cdot\mathbf{\widehat
{\mathbf{p}}}+m_{0}c^{2}\Gamma_{2}\right)  \Psi\left(  \mathbf{r}\right)  ,
\label{801}%
\end{equation}
i.e.,
\begin{equation}
\left\{
\begin{array}
[c]{c}%
2\left(  E_{\pm}-eV\left(  r\right)  \right)  \phi\left(  \mathbf{r}\right)
=-c\mathbf{\sigma}\cdot\mathbf{\widehat{\mathbf{p}}}\chi\left(  \mathbf{r}%
\right)  +m_{0}c^{2}\left(  \phi\left(  \mathbf{r}\right)  -i\chi\left(
\mathbf{r}\right)  \right)  ,\\
\qquad\qquad\qquad\quad\ \ 0=-c\mathbf{\sigma}\cdot\mathbf{\widehat
{\mathbf{p}}}\phi\left(  \mathbf{r}\right)  +m_{0}c^{2}\left(  i\phi\left(
\mathbf{r}\right)  -\chi\left(  \mathbf{r}\right)  \right)  .
\end{array}
\right.  \label{81}%
\end{equation}
The second equation (\ref{81}) gives the spinor $\chi\left(  \mathbf{r}%
\right)  $ in terms of the spinor $\phi\left(  \mathbf{r}\right)  $:
\begin{equation}
\chi\left(  \mathbf{r}\right)  =-\left(  im_{0}c^{2}+c\mathbf{\sigma}%
\cdot\mathbf{\widehat{\mathbf{p}}}\right)  \phi\left(  \mathbf{r}\right)
/m_{0}c^{2}. \label{82}%
\end{equation}
Inserting $\chi\left(  \mathbf{r}\right)  $ into the first Eq.(\ref{81}), we
finally arrive to
\begin{equation}
\widehat{H}\phi\left(  \mathbf{r,}t\right)  =\left(  m_{0}c^{2}+eV\left(
r\right)  +\frac{\left(  \mathbf{\sigma}\cdot\mathbf{\widehat{\mathbf{p}}%
}\right)  ^{2}}{2m_{0}}\right)  \phi\left(  \mathbf{r},t\right)  =i\hbar
\frac{\partial}{\partial t}\phi\left(  \mathbf{r},t\right)  , \label{100}%
\end{equation}
which is the `Pauli' equation for this system, wherein negative-energy
eigenstates have been included, up to a minus (-) sign on the right hand side
of Eq.(\ref{100}). Thus a general solution $\phi\left(  \mathbf{r},t\right)  $
to Eq.(\ref{100}) is a linear combination of positive and negative-eigenstates
$\phi\left(  \mathbf{r}\right)  \exp\left(  -\frac{i}{\hbar}E_{+}t\right)  $
and $\phi\left(  \mathbf{r}\right)  \exp\left(  -\frac{i}{\hbar}E_{-}t\right)
$.

Of course, it is possible to consider a general minimal interaction in
Eq.(\ref{20}):
\begin{equation}
\widehat{H}=c\widehat{p}_{0}\rightarrow c\widehat{p}_{0}-eV(\mathbf{r,}%
t),\qquad\widehat{\mathbf{p}}\rightarrow\widehat{\mathbf{\Pi}}=\widehat
{\mathbf{p}}-\frac{e}{c}\mathbf{A}(\mathbf{r,}t). \label{102}%
\end{equation}
Thereafter, we go along the same steps as above. Alternatively, one can use
either Eq.(\ref{32}) or Eq.(\ref{100}) to introduce the remaining piece of
minimal coupling.

It is interesting to note that an \textit{ubiquitous} $m_{0}c^{2}%
\times\widehat{\mathcal{O}}$ additive term is present in equations (\ref{20}),
(\ref{801}) and (\ref{100}), amongst some others throughout the text. This
fact might be paradoxically interpreted as follows: we need to introduce a
\textit{relativistic} concept (rest energy$\ =E_{0}=m_{0}c^{2}$) in order to
be able to build a strictly \textit{nonrelativistic} ($\left\vert
\mathbf{v}\right\vert \ll c$) wave equation.

This work was supported by Direcci\'{o}n de Investigaci\'{o}n, Universidad de
Concepci\'{o}n, through grant P.I. 207.011.046-1.0.

\end{document}